\documentclass[reprint,superscriptaddress,amsmath,amssymb,aps,floatfix,longbibliography]{revtex4-1}
\usepackage{graphicx}
\usepackage{color}

\usepackage{dcolumn}
\usepackage{upgreek}
\usepackage{gensymb}
\usepackage{float}
\usepackage{bm}
\usepackage{xcolor}

\usepackage{xr}
\makeatletter
\newcommand*{\addFileDependency}[1]{
  \typeout{(#1)}
  \@addtofilelist{#1}
  \IfFileExists{#1}{}{\typeout{No file #1.}}
}
\makeatother

\newcommand*{\myexternaldocument}[1]{
    \externaldocument{#1}
    \addFileDependency{#1.tex}
    \addFileDependency{#1.aux}
}

\myexternaldocument{supps_083120}

\begin{document}
	
\title{Giant inverse Faraday effect in Dirac semimetals}

\author{Masashi Kawaguchi}
\thanks{Equally contributed}
\affiliation{Department of Physics, The University of Tokyo, Tokyo 113-0033, Japan}

\author{Hana Hirose}
\thanks{Equally contributed}
\affiliation{Department of Physics, The University of Tokyo, Tokyo 113-0033, Japan}

\author{Zhendong Chi}
\affiliation{Department of Physics, The University of Tokyo, Tokyo 113-0033, Japan}

\author{Yong-Chang Lau}
\affiliation{Department of Physics, The University of Tokyo, Tokyo 113-0033, Japan}

\author{Frank Freimuth}
\affiliation{Peter Gr\"{u}nberg Institut and Institute for Advanced Simulation, Forschungszentrum J\"{u}lich and JARA, 52425 J\"{u}lich, Germany}

\author{Masamitsu Hayashi}
\affiliation{Department of Physics, The University of Tokyo, Tokyo 113-0033, Japan}

\newif\iffigure
\figurefalse
\figuretrue

\date{\today}

\begin{abstract}
We have studied helicity dependent photocurrent (HDP) in Bi-based Dirac semimetal thin films.
HDP increases with film thickness before it saturates, changes its sign when the majority carrier type is changed from electrons to holes and takes a sharp peak when the Fermi level lies near the charge neutrality point. 
These results suggest that irradiation of circularly polarized light to Dirac semimetals induces an effective magnetic field that aligns the carrier spin along the light spin angular momentum and generates a spin current along the film normal.
The effective magnetic field is estimated to be orders of magnitude larger than that caused by the inverse Faraday effect (IFE) in typical transition metals.
We consider the small effective mass and the large $g$-factor, characteristics of Dirac semimetals with strong spin orbit coupling, are responsible for the giant IFE, opening pathways to develop systems with strong light-spin coupling.
\end{abstract}

\maketitle



Conservation of the spin angular momentum plays an essential role in the interaction between light and electron spin in solids.
The optical selection rule, \textit{i.e.}, the Fermi's golden rule, sets the transition rate of electrons from the ground state to the excited state upon light irradiation.
Exploiting the selection rule, one can excite spin polarized carriers in semiconductors\cite{zutic2002prl,ganichev2004prl,ando2010apl,endres2013ncomm} or in systems with spin-momentum locked bands\cite{mciver2012nnano,yuan2014nnano,okada2016prb,pan2017ncomm,ma2017nphys,puebla2019prl}.
It is generally understood that the efficiency to excite such carriers is the largest when the light energy is close to the energy band gap of the system.

Significant light induced effects have also been observed in metals, where there is no band gap.
For example, application of ultrashort polarized laser pulses to magnetic thin films allows manipulation of the magnetization\cite{beaurepaire1996prl,kimel2005nature,satoh2010prl} and can lead to light induced magnetization switching\cite{stanciu2007prl,lambert2014science}.
In many cases, the orientation of the magnetization is defined by the helicity of circularly polarized light.
Recent studies have shown that irradiation of ultrashort laser pulses to ferromagnetic metal/heavy metal bilayers results in emission of electromagnetic waves in the THz range\cite{kampfrath2013nnano,huisman2016nnano,seifert2016nphoton}.
In this case, linearly polarized light is typically used to generate THz signals.
Although the underlying mechanism of all optical switching and THz emission is under debate, two different processes are considered to contribute to the effects\cite{berritta2016prl,qaiumzadeh2016prb,freimuth2016prb,tokman2020prb}. 
The optical spin transfer torque is a direct transfer of spin angular momentum from light to electrons\cite{choi2015nphys}, while the inverse Faraday effect (IFE) can be considered equivalent to applying an effective magnetic field associated with circularly polarized light\cite{kimel2005nature,stanciu2007prl,lambert2014science,john2017scirep}.
As the light intensity required to observe these effects are extremely large, means to increase their efficiencies are being sought.

Here we show, from helicity dependent photocurrent measurements, that irradiation of circularly polarized light to Bi-based Dirac semimetals\cite{murakami2006prl,li2008science,fuseya2009prl} induces a giant effective magnetic field along the light spin angular momentum.
The experimental results suggest that the effective magnetic field peaks at the Dirac point with its maximum being significantly larger than that of typical transition metals induced by the IFE.
We consider the giant IFE is due to the unique characteristics of carriers in Dirac semimetals.
Semimetal thin films are deposited on SiO$_2$ substrates using RF magnetron sputtering.
Here we show representative results from Bi, Bi$_{1-x}$Sb$_x$ alloy, Sn- or Te-doped Bi thin films with a thickness of $t$ (see supplementary material for the details of sample preparation and device characterization).
A seed layer of 0.5 Ta/2 Cu (thickness in nm) is inserted for the Bi$_{1-x}$Sb$_x$ alloy film.
The influence of the seed layer on the photocurrent will be discussed elsewhere.
A metal shadow mask is inserted between the substrate and the sputtering target to form a wire\cite{hirose2018apl}.

\begin{figure}[t]
 \begin{center}
  \includegraphics[scale=0.7]{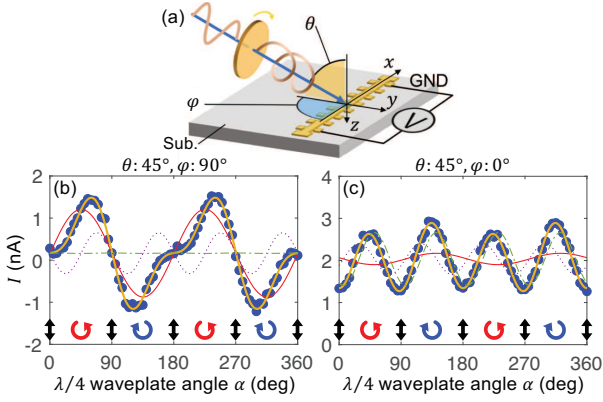}
  \caption{
  (a) Schematic illustration of the experimental setup and definition of the coordinate axis. The yellow line on the substrate represents the wire made of the film. Light is irradiated from an angle ($\theta$, $\phi$) as defined in the image. 
  (b,c) Bottom panels: The quarter wave plate optical axis angle ($\alpha$) dependence of the photocurrent ($I$) for Bi thin film with $t\sim70$ nm. $\phi\sim90^\circ$ (b) and  $\phi\sim 0^\circ$ (c). $\theta$ is fixed to $\sim45^\circ$. The solid circles represent experimental data, the orange solid line shows fit to the data with Eq.~(\ref{eq:fitting}). The red solid, purple dotted and green dashed lines show contributions from the $C$, $L_1$, and $L_2$ terms, respectively. 
  (b,c) Data are obtained using $\lambda=405$ nm and $P\sim2.5$ mW.
  }
  \label{fig:1}
 \end{center}
\end{figure}

Schematic illustration of the experimental setup and the coordinate axis are shown in Fig.~\ref{fig:1}(a).
We measure the photocurrent of the wire made of semimetal thin films while illuminating light through a quarter wave plate.
A continuous wave semiconductor laser light with wavelength $\lambda$ and power $P$ is used as the light source. 
$\theta$ and $\phi$ are the polar and azimuthal angles of the light incidence.
The quarter wave plate is rotated to change the light helicity. The incident light is linearly polarized when $\alpha=0^{\circ}, 90^{\circ}, 180^{\circ}, 270^{\circ}$ and circularly polarized when $\alpha=45^{\circ}, 225^{\circ}$ (left handed) and $135^{\circ}, 315^{\circ}$ (right handed). The left (right) handed circularly polarized light has an angular momentum of $\hbar$ against (along) the light propagation direction, where $\hbar$ is the reduced Planck constant.

Representative results of the photocurrent are plotted in Figs.~\ref{fig:1}(b) and \ref{fig:1}(c) as a function of $\alpha$ for Bi thin film ($t\sim$70 nm).
The incident angle $\theta$ is fixed to $\sim45^{\circ}$ and $\phi\sim90^\circ$ for Fig.~\ref{fig:1}(b) and $\phi\sim0^{\circ}$ for Fig.~\ref{fig:1}(c). A large helicity dependent photocurrent (period of 180$^\circ$) is found when $\phi\sim90^\circ$ whereas the effect is negligibly small when $\phi\sim 0^\circ$. 
The $\alpha$ dependence of the photocurrent is fitted with the following function\cite{mciver2012nnano,okada2016prb}:
\begin{equation}
\begin{aligned}
    I=&C\sin{2(\alpha + \alpha_0)}\\
    &+L_1\sin{4(\alpha + \alpha_0)}+L_2\cos{4(\alpha + \alpha_0)}+I_\mathrm{0}
\label{eq:fitting}
\end{aligned}
\end{equation}
where $C$ represents the difference in the photocurrent when left and right handed polarized light are illuminated. 
$L_1$ and $L_2$ are the changes in the photocurrent under illumination of circularly and linearly polarized light. 
The last term $I_\mathrm{0}$ is a constant term that is independent of $\alpha$.
$\alpha_0$ is an offset angle associated with the optical setup and is $\sim-1^{\circ}$ here.
The fitting results are shown using the orange solid line in Figs.~\ref{fig:1}(b) and \ref{fig:1}(c).
Large contribution from the $C$ term (red solid line) is found for $\phi\sim90^\circ$ whereas the $L_2$ term (green dashed line) dominates for $\phi\sim 0^\circ$.
We focus on the characteristics of $C$, obtained using $\phi\sim90^\circ$, hereafter.

\begin{figure}[t]
\centering
\includegraphics[scale=0.7]{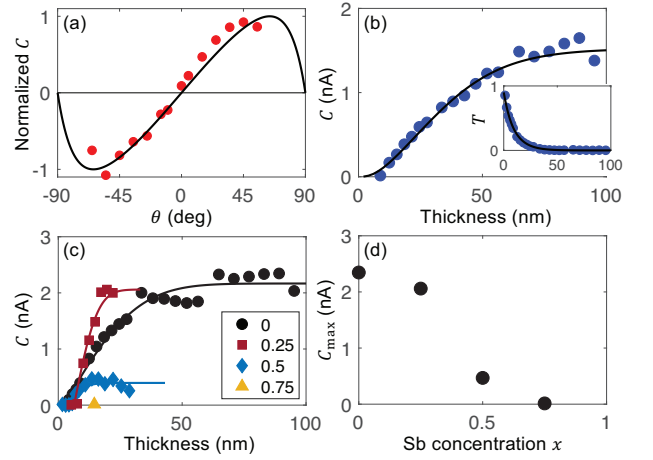}
\caption{
(a) The light incident angle ($\theta$) dependence of the helicity dependent photocurrent $C$ with $\phi\sim90^\circ$ for $\sim30$ nm thick Bi thin film. $C$ is normalized by its maximum value obtained at $\theta\sim\pm45^{\circ}$. The solid line shows the calculated $\theta$ dependence of $P_\mathrm{a}\sin{\theta}$. (b) The Bi layer thickness ($t$) dependence of $C$ for Bi thin film. The inset shows the corresponding $t$ dependence of the transmission coefficient ($T$). (c) $C$ vs. $t$ for Cu-seeded Bi$_{1-x}$Sb$_x$ thin films. The Sb concentration ($x$) is denoted in the box. (b,c) The lines are guides to the eye. Data are obtained using $\theta\sim45^\circ$, $\phi\sim90^\circ$. (d) $x$ dependence of maximum $C$ ($C_\mathrm{max}$) obtained from (c). (a-d) Laser with $\lambda=405$ nm and $P\sim2.5$ mW is used to obtain the data.} 
\label{fig:2}
\end{figure}

The incident angle $\theta$ dependence of $C$ is shown in Fig.~\ref{fig:2}(a).
$C$ is normalized by its maximum absolute value obtained at $\theta\sim\pm45^{\circ}$.
The $\theta$ dependence of the normalized $C$ can be accounted for with a functional form $P_\mathrm{a}\sin{\theta}$, which is shown by the solid line in Fig.~\ref{fig:2}(a).
$P_\mathrm{a}$ is the absorbance of the circularly polarized light calculated for a Bi thin film in contact with air (the optical constants of Bi are obtained experimentally that are consistent with previous report\cite{werner2009jpcrd}). 
$\sin{\theta}$ represents the in-plane component of the light spin angular momentum.
These results show that the in-plane component of the light spin angular momentum plays an essential role in setting the helicity dependent photocurrent. 
From hereon, we discuss results obtained using $\theta\sim45^\circ$.

The Bi layer thickness ($t$) dependence of $C$ is shown in Fig.~\ref{fig:2}(b).
We find $C$ increases with $t$ and tends to saturate at larger $t$.
Such thickness dependence suggests that a large part of the photocurrent is generated within the bulk of the film.
To highlight the unique characteristics of Bi, we have studied photocurrent in Bi$_{1-x}$Sb$_x$ thin films.
Although Bi-rich Bi$_{1-x}$Sb$_x$ alloy is known as a three-dimensional topological insulator\cite{hsieh2008nature}, we consider it unlikely that the films grown by sputtering host topological surface states given the polycrystalline structure of the films\cite{chi2020sciadv}. 
The $t$ dependence of $C$ for Bi$_{1-x}$Sb$_x$ thin films are presented in Fig.~\ref{fig:2}(c).  
In all structures except for $x\sim0.75$, whose $C$ is nearly zero, $C$ increases with $t$ until it saturates. 
The maximum value of $C$ for each structure is defined as $C_\mathrm{max}$: $C_\mathrm{max}$ is plotted as a function of $x$ in Fig.~\ref{fig:2}(d).
$C_\mathrm{max}$ decreases with increasing $x$.
Interestingly, the $x$ dependence of $C_\mathrm{max}$ resembles that of the spin Hall conductivity of Bi$_{1-x}$Sb$_x$ alloy\cite{chi2020sciadv,sahin2015prl}.
These results thus suggest that spin current plays an important role in the generation of helicity dependent photocurrent. 

\begin{figure}[h]
 \begin{center}
  \includegraphics[scale=0.7]{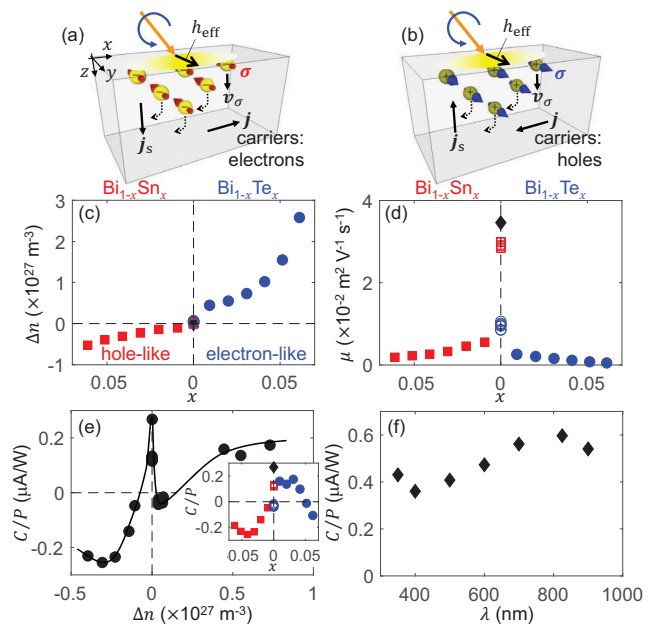}
  \caption{
  (a,b) Schematic illustration of light induced spin density. The in-plane component of the spin angular momentum ($\bm{\sigma}$) of the electrons (a) and holes (b) is sketched using the arrows that thread the yellow spheres when a right-handed circularly polarized light is irradiated to the film. The gradient in the spin density causes carriers to diffuse along the film normal with velocity ($\bm{v}_\sigma$) and generates a spin current $\bm{j}_\mathrm{s}$. $\bm{j}_\mathrm{s}$ is converted to charge current ($\bm{j}$) via the ISHE. $h_{\mathrm{eff}}$ represents the $y$-component of the light induced effective magnetic field ($\bm{h}_\mathrm{eff}$). (c,d) Doping concentration ($x$) dependence of the difference of the electron density ($n_e$) and hole density ($n_h$), $\Delta n \equiv n_e - n_h$ (c), and the mobility ($\mu$) (d) of Sn-doped Bi (Bi$_{1-x}$Sn$_x$) and Te-doped Bi (Bi$_{1-x}$Te$_x$) thin films with $t\sim65-70$ nm. Red squares: Bi$_{1-x}$Sn$_x$, blue circles: Bi$_{1-x}$Te$_x$, black diamond: Bi. The open symbols represent results from thin films with lightly doped Bi. (e) $C/P$ plotted against $\Delta n$ for all the structures shown in (c,d). The solid line is guide to the eye. The inset shows the $x$ dependence of $C/P$. Laser with $\lambda=405$ nm and $P\sim2.5$ mW is used to obtain the data. (f) Wavelength dependence of $C/P$ for Bi thin film with $t\sim65$ nm ($P\sim2.5$ mW). (e,f) Data are obtained using $\theta\sim45^\circ$, $\phi\sim90^\circ$. }
  \label{fig:3}
 \end{center}
\end{figure}

Based on the results presented in Fig.~\ref{fig:2}, we consider the generation of helicity dependent photocurrent involves two processes: (1) light induced generation of spin density and (2) conversion of the spin density to charge current. 
A schematic illustration of the two processes is sketched in Fig.~\ref{fig:3}(a).
In the first process, a circularly polarized light induces spin density, an imbalance in the electron population with spin angular momentum pointing parallel and antiparallel to the light propagation direction.
Assuming that the spin density scales with the light absorbance, the finite penetration depth ($\lambda_\mathrm{p}$) of light into the semimetal causes a gradient in the spin density along the film normal. 
See Fig.~\ref{fig:2}(b) inset for a representative thickness dependence of the transmission coefficient ($T$) of Bi thin film, which is roughly proportional to the light absorbance.
We find $\lambda_\mathrm{p} \sim 10$ nm for Bi.
The spin density gradient induces a spin current along the film normal, which in the second process is converted to charge current via the inverse spin Hall effect (ISHE)\cite{saitoh2006apl}.
Under the ISHE, the charge current along $i$ ($j_i$) is expressed as 
\begin{equation}
\begin{aligned}
	j_i = \theta_\mathrm{SH} e n ( \bm{v}_\sigma \times \bm{\sigma} )_i,
\label{eq:ishe}
\end{aligned}
\end{equation}
where $\theta_\mathrm{SH}$ is the spin Hall angle, $e$ is the elementary charge ($e>0$), $n$ is the carrier density, $\bm{\sigma}$ is a vector that represents the spin angular momentum of the carriers and $\bm{v}_\sigma$ is the velocity of the carriers with spin $\bm{\sigma}$. $\bm{v}_\sigma$ is defined as $\displaystyle{\bm{v}_\sigma = - \mu \nabla ( \frac{u_\sigma}{2e}} )$, where $\mu$ is the carrier mobility and $u_\sigma$ is the chemical potential of carriers with spin $\bm{\sigma}$.
For a reason that will be discussed later, we assume that $\bm{\sigma}$ points along, or against, the light propagation direction ($\bm{l}$).
Since $\bm{v}_\sigma$ is parallel to the light intensity gradient (along $z$ in Fig.~\ref{fig:3}(a)), $|j_x|$ scales with the in-plane component of $\bm{l}$ and takes a maximum when the $y$-component of $\bm{l}$ is the largest (i.e. when $\phi=90, 270^\circ$).
These features are consistent with the $\theta$ and $\phi$ dependences of $C$.

To clarify how the electronic structure of Bi influences the generation of helicity dependent photocurrent, we have doped Bi with Sn and Te to vary the carrier concentration and change the position of the Fermi level with respect to the Dirac point\cite{liu1995prb,fuseya2009prl}.
Figure~\ref{fig:3}(c) shows the nominal doping concentration ($x$) dependence of the difference of the electron density ($n_e$) and hole density ($n_h$), $\Delta n \equiv n_e - n_h$, of Sn-doped Bi (Bi$_{1-x}$Sn$_x$) and Te-doped Bi (Bi$_{1-x}$Te$_x$) thin films.
Note that for pure Bi, the number of electrons and holes are equal due to the charge neutrality condition and the carrier concentration ($n = n_e + n_h$) of the films is $\sim1.1-1.2 \times 10^{25}$ m$^{-3}$.
The majority carrier type of Sn- and Te-doped Bi is holes and electrons, respectively.
For both systems, $|\Delta n|$ decreases with decreasing $x$.
In contrast, the mobility ($\mu$), shown in Fig.~\ref{fig:3}(d), sharply increases as $x$ is reduced, taking a peak at $x\sim0$. 


The inset of Fig.~\ref{fig:3}(e) shows the $x$ dependence of $C$ divided by the laser power $P$. 
We find that the sign of $C$ reverses as the majority carrier type is changed from electrons to holes.
The sign change in $C$ across $x\sim0$ can be understood if we assume the inverse Faraday effect (IFE) is responsible for the generation of light induced spin density.
In this process, circularly polarized light acts as an effective magnetic field ($\bm{h}_\mathrm{eff}$), which is parallel (antiparallel) to $\bm{l}$ for right (left)-handed circularly polarized light.
Application of $\bm{h}_\mathrm{eff}$ will align the magnetic moment ($\bm{m}$) of the carriers along $\bm{h}_\mathrm{eff}$.
Consequently, the spin angular momentum ($\bm{\sigma}$) of the electrons will point opposite to $\bm{h}_\mathrm{eff}$ whereas that of the holes will be parallel to $\bm{h}_\mathrm{eff}$. 
Theoretical study shows that the sign of $\theta_\mathrm{SH}$ for electron and hole doped Bi is the same\cite{sahin2015prl}.
Thus change in the majority carrier from electrons to holes causes $\bm{\sigma}$ in Eq.~(\ref{eq:ishe}) to reverse its direction, resulting in $C$ with opposite sign (see Figs.~\ref{fig:3}(a) and \ref{fig:3}(b) for a schematic illustration).
(For the Te-doped Bi thin films, the sign of $C$ changes at $x\sim 0.04$, which coincides with $x$ where a rapid increase in the carrier concentration is found. 
We consider the system becomes degenerate and contributions from other bands influence $C$ in this regime ($x \geq 0.04$).) 



From the magnitude of $C$, we first provide a rough estimate of the light induced spin density per unit volume ($n_\mathrm{s}$).
$n_\mathrm{s}$ can be obtained from $N_\mathrm{F}$, the density of states at the Fermi level, and $u_\mathrm{s}$, the difference in the chemical potential of carriers with $\bm{\sigma}$ parallel and antiparallel to the in-plane component of $\bm{l}$, via the relation $n_\mathrm{s} = N_\mathrm{F} u_\mathrm{s}$.
As a first order estimation, we assume that $u_\mathrm{s}$ varies linearly from $u_\mathrm{s,max}$ at the surface of the semimetal to 0 over a distance ($\lambda_\mathrm{s,p}$) set either by the larger of the spin diffusion length ($\lambda_\mathrm{s}$) and the light penetration length ($\lambda_\mathrm{p}$).
Consequently, a spin current of $\displaystyle{j_\mathrm{s} = \frac{\sigma_{xx} u_\mathrm{s,max}}{2 e \lambda_\mathrm{s,p}}}$\cite{chen2013prb} flows along the film normal, where $\sigma_{xx}$ is the electrical conductivity of the semimetal.
The resulting charge current due to the ISHE (Eq.~(\ref{eq:ishe})), equivalent to $C$ obtained in the experiments, is calculated by multiplying $j_\mathrm{s}$ with $\lambda_\mathrm{s,p} w \theta_\mathrm{SH}$, where $w$ is the width of the wire.
The maximum $n_\mathrm{s}$ induced at the surface of the semimetal layer ($n_\mathrm{s,max}$) therefore reads
\begin{equation}
\begin{aligned}
	n_\mathrm{s,max} \sim \frac{2 e N_\mathrm{F} C}{\sigma_{xx} \theta_\mathrm{SH} w}.
\label{eq:ns}
\end{aligned}
\end{equation}
Substituting $C \sim 1$ nA, $w \sim 0.4$ mm, $\sigma_{xx} \sim 1.3 \times 10^5$ ($\Omega^{-1}$ m$^{-1}$) and $N_\mathrm{F} \sim 1.4 \times 10^{28}$ (eV$^{-1}$ m$^{-3}$)\cite{gonze1990prb} for Bi, we find $\frac{n_\mathrm{s,max}}{P/S} \sim \frac{4.5 \times 10^{13}}{\theta_\mathrm{SH}}$ (W$^{-1}$ m$^{-1}$), where $S$ is the laser irradiation area.
Even with a spin Hall angle of $\sim1$\cite{chi2020sciadv}, $\frac{n_\mathrm{s,max}}{P/S}$ is three to four orders of magnitude larger than that of typical transition metals\cite{berritta2016prl,freimuth2016prb,huisman2016nnano}.

\begin{figure}[t]
 \begin{center}
    \includegraphics[scale=0.7]{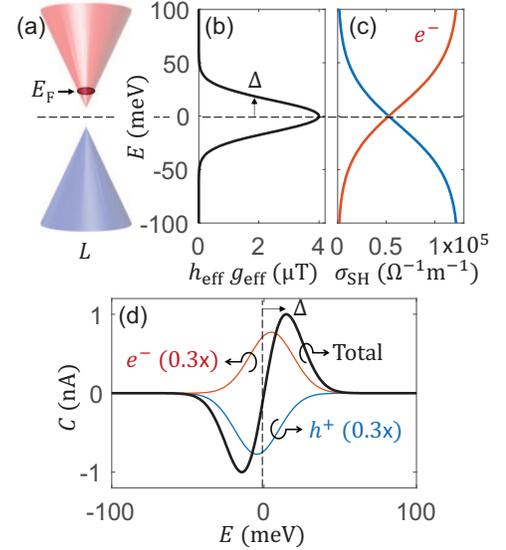}
  \caption{
  (a) Schematic illustration of the electronic band structure of Bi at the $L$ valley. $E_\mathrm{F}$ denotes the Fermi level of Bi. (b-d) Fermi level position dependence of the product of the light induced effective magnetic field ($h_\mathrm{eff}$) and the $g$-factor ($g_\mathrm{eff}$), $h_\mathrm{eff} g_\mathrm{eff}$ (b), spin Hall conductivity $\sigma_\mathrm{SH}$ (c) and the helicity dependent photocurrent $C$ (d) calculated using Eqs.~(\ref{eq:Call}) and (\ref{eq:CLTeh}). Red and blue solid lines in (c,d) indicate contribution from electrons and holes in the $L$ valley. $\Delta$ indicates the width of the $h_\mathrm{eff} g_\mathrm{eff}$ peak.}
  \label{fig:4}
 \end{center}
\end{figure}

The $\Delta n$ dependence of $C$ is presented in Fig.~\ref{fig:3}(e).
Clearly, $C$ takes a maximum at $\Delta n \sim 0$.
As the electron and hole densities are the same for Bi, however, $C$ should be zero at $\Delta n \sim 0$ if the scenario sketched in Figs.~\ref{fig:3}(a,b) applies.
To account for these results, we develop a toy model (a rigid band model is assumed under Sn/Te doping).
As illustrated in Fig.~\ref{fig:4}(a), the Fermi level of Bi ($\Delta n \sim 0$) lies above the Dirac point at the $L$ valley.
The $L$ valley supplies electrons that contribute to transport whereas the $T$ valley provides holes\cite{liu1995prb,fuseya2015jpsj}.
Assuming that carriers in each valley contribute to $C$ independently, we obtain $C$ by summing all contributions (and taking into account that the signs of $C$ due to electrons and holes are opposite):
\begin{equation}
\begin{aligned}
	C \sim C_{L,e^-} - C_{L,h^+} - C_{T,h^+},
\label{eq:Call}
\end{aligned}
\end{equation}
where $C_{v, c}$ represents $C$ due to carrier type $c$ ($e^-$: electrons. $h^+$: holes) at valley $v$ ($L$ or $T$). 
(Note that the $T$ valley only accommodates holes).
We assume that the spin density is equal to the product of the Zeeman energy and $N_\mathrm{F}$ for each valley, \textit{i.e.}, $n_\mathrm{s,max} \sim \mu_\mathrm{B} g_\mathrm{eff} h_\mathrm{eff} N_\mathrm{F}$, where $\mu_\mathrm{B}$, $g_\mathrm{eff}$ and $h_\mathrm{eff}$ are the Bohr magneton, the effective $g$-factor and the $y$-component of $\bm{h}_\mathrm{eff}$, respectively.
Using Eq.~(\ref{eq:ns}), $C_{v,c}$ can be expressed as
\begin{equation}
\begin{gathered}
	C_{v, c} \sim \frac{\mu_\mathrm{B} w}{2 e} \big(h_\mathrm{eff} g_\mathrm{eff} \sigma_\mathrm{SH} \big)\big|_{v,c},
\label{eq:CLTeh}
\end{gathered}
\end{equation}
where $\sigma_\mathrm{SH} \equiv \sigma_{xx} \theta_\mathrm{SH}$ is the spin Hall conductivity.

Theoretical calculations suggest that $\sigma_\mathrm{SH}$ is larger for the electrons in the $L$ valley than that of the holes in the $T$ valley\cite{sahin2015prl,fuseya2015jpsj}.
Such difference breaks the electron-hole symmetry of $C$ at $\Delta n \sim 0$: we neglect contribution from $C_{T,h^+}$ hereafter.
Calculations also indicate that $\sigma_\mathrm{SH}$ of the carriers in the $L$ valley is nearly constant across the Dirac point\cite{sahin2015prl}.
We thus assume $\sigma_\mathrm{SH}$ is constant for the majority carriers and scales with the carrier density for the minority carriers, as shown in Fig.~\ref{fig:4}(c) (see supplementary material for the details).
The calculated $\Delta n$ dependence of $C$ is presented in Fig.~\ref{fig:4}(d).
We find the experimental results are best described if $h_\mathrm{eff} g_\mathrm{eff}$ takes a peak at the Dirac point and its width is close to the energy difference between the Dirac point and the Fermi level of Bi, as sketched in Fig.~\ref{fig:4}(b).
Note that the peak width of $h_\mathrm{eff} g_\mathrm{eff}$, defined as $\Delta$ in Fig.~\ref{fig:4}(b), roughly defines the position of $C$ maximum; see Fig.~\ref{fig:4}(d).


In the calculation, we adjust the peak value of $h_\mathrm{eff} g_\mathrm{eff}$ so that the maximum $C$ at $\Delta n \sim 0$ is $\sim 1$ nA ($\theta_\mathrm{SH} \sim 1$ is assumed), in accordance with the experiments.
Past studies have shown that the electrons in the $L$ valley possess large $g_\mathrm{eff}$, of the order of $\sim$100 to $\sim$1000\cite{fuseya2015prl} owing to its small effective mass and large spin orbit coupling.
Even with $g_\mathrm{eff} \sim 1000$, which is unlikely given the poly-crystal texture of the films used here, the estimated $h_\mathrm{eff}$ ($\sim 4 \times 10^{-9}$ T under light intensity of $\sim 1.3 \times 10^4$ W/m$^2$) is one to two orders of magnitude larger than that of typical transition metals caused by the IFE\cite{berritta2016prl,freimuth2016prb,huisman2016nnano}.
Thus the large $h_\mathrm{eff} g_\mathrm{eff}$ at the Dirac point is likely due to contributions from both $h_\mathrm{eff}$ and $g_\mathrm{eff}$.
$g_\mathrm{eff}$ may follow the $\Delta n$ dependence of mobility $\mu$ (Fig.~\ref{fig:3}(d)) via changes in the carrier effective mass.
Similarly, $h_\mathrm{eff}$ may scale with the inverse of the carrier effective mass, as recent calculations suggest\cite{tokman2020prb}.
We thus consider the unique characteristics of the Dirac semimetals with strong spin orbit coupling, \textit{i.e.}, small effective mass and large $g$-factor, cause the giant IFE in Bi.
Note that in the experiments, we find a dip in $C$ right above $\Delta n \sim 0$, which cannot be accounted for with this model. 
To fully describe the experimental results, we infer that influence of carrier density on $C$ must also be taken into account.

Theoretical calculations show that the strength of IFE increases with increasing light wavelength ($\lambda$)\cite{berritta2016prl,freimuth2016prb}.
Figure.~\ref{fig:3}(f) shows the $\lambda$ dependence of $C$ for Bi thin film ($\Delta n \sim 0$).
As the light power ($P$) depends on $\lambda$ (see supplementary material), $C$ is normalized by $P$.
Although the range of $\lambda$ studied here is limited, $C/P$ tends to increase with increasing $\lambda$. 
It remains to be seen whether $C$ will further increase when the light energy becomes close to band gap of Bi.
See supplementary material for discussion on contributions to the photocurrent from other sources.


In summary, we have shown that a giant light induced effective magnetic field emerges in Dirac semimetals.
A spin density that scales with the light intensity is created by the IFE, causing a spin current along the film normal.
The spin current is converted to charge current via the ISHE, giving rise to HDP.
The carrier transport characteristics of Dirac semimetals with strong spin orbit coupling are responsible for the giant IFE.
%
These results demonstrate the unique optical response of Dirac semimetals that can be exploited to develop systems with strong light-spin coupling.

\begin{acknowledgments}
This work was partly supported by JST CREST (JPMJCR19T3), JSPS Grant-in-Aids (JP15H05702,JP16H03853), Yamada Science Foundation and the Center of Spintronics Research Network of Japan. Y.-C.L. is supported by JSPS International Fellowship for Research in Japan (JP17F17064). H.H. and Z.C. acknowledge financial support from Materials Education program for the future leaders in Research, Industry, and Technology (MERIT). 
\end{acknowledgments}

\bibliography{ref_060220.bib}


\end{document}